\documentclass[aps,notitlepage,preprintnumbers,onecolumn,amsmath,amssymb,floatfix,superscriptaddress,pra]{revtex4-1}

\pdfoutput=1
\usepackage[utf8]{inputenc}
\usepackage{amsmath}
\usepackage{amsfonts}
\usepackage{braket}
\usepackage{graphicx}
\usepackage{empheq}
\usepackage{amsthm}

\def\be{\begin{equation}}
	\def\ee{\end{equation}}
\def\bea{\begin{eqnarray}}
	\def\eea{\end{eqnarray}}

\usepackage{tikz}
\usetikzlibrary{patterns}
\usetikzlibrary{snakes}
\usetikzlibrary{decorations.pathreplacing}
\usetikzlibrary{plotmarks}
\definecolor{myred}{RGB}{232,102,102}
\definecolor{myblue}{RGB}{187,187,255}
\definecolor{mygreen}{RGB}{34,139,34}

\usetikzlibrary{shapes.misc}
\usetikzlibrary{decorations.pathmorphing}

\tikzset{cross/.style={cross out, draw=black, minimum size=2*(#1-\pgflinewidth), inner sep=0pt, outer sep=0pt},
	cross/.default={1pt}}

\usepackage[colorlinks=true,linkcolor=blue, citecolor=blue, bookmarks]{hyperref}

\begin{document}
	
	\title{Thermodynamic symmetry resolved entanglement entropies in integrable systems}
	\date{\today}
	\author{Lorenzo Piroli}
	\affiliation{Philippe Meyer Institute, Physics Department, \'Ecole Normale Sup\'erieure (ENS), Universit\'e PSL, 24 rue Lhomond, F-75231 Paris, France}
	\author{Eric Vernier}
	\affiliation{CNRS \& Université Paris Cité, Laboratoire de Probabilités, Statistique et Modélisation, F-75013 Paris, France
	}
	\author{Mario Collura}
	\affiliation{SISSA, via Bonomea 265, 34136 Trieste, Italy}
	\affiliation{INFN Sezione di Trieste, via Bonomea 265, 34136 Trieste, Italy}
	\author{Pasquale Calabrese}
	\affiliation{SISSA, via Bonomea 265, 34136 Trieste, Italy}
	\affiliation{INFN Sezione di Trieste, via Bonomea 265, 34136 Trieste, Italy}
	\affiliation{The Abdus Salam International Center for Theoretical Physics, Strada Costiera 11, 34151 Trieste, Italy}
	\begin{abstract}
	We develop a general approach to compute the symmetry-resolved R\'enyi and von Neumann entanglement entropies (SREE) of thermodynamic macrostates in interacting integrable systems. Our method is based on a combination of the thermodynamic Bethe ansatz and the G\"artner-Ellis theorem from large deviation theory. We derive an explicit simple formula for the von Neumann SREE, which we show to coincide with the thermodynamic Yang-Yang entropy of an effective macrostate determined by the charge sector. Focusing on the XXZ Heisenberg spin chain, we test our result against iTEBD calculations for thermal states, finding good agreement. As an application, we provide analytic predictions for the asymptotic value of the SREE following a quantum quench. 
	\end{abstract}
	\maketitle

\section{Introduction}

In the past two decades, the study of different measures of quantum entanglement, such as the von Neumann and R\'enyi entropies~\cite{nielsen2002quantum},  have revolutionized our understanding of many-body systems, with ramifications ranging from high energy physics to the theory of critical and non-equilibrium phenomena~\cite{amico2008entanglement,calabrese2009entanglement,eisert2010colloquium,laflorencie2016quantum,h-19}. Entanglement entropies play in particular a crucial role in the emergence of thermodynamics in isolated many-body systems out of equilibrium~\cite{santos2011entropy,ckc-13,deutsch2013microscopit,alba2017entanglement}, as recently demonstrated by beautiful experiments in cold-atom settings~\cite{kaufman2016quantum,brydges2019probing,elben2020mixed}. For example, in the framework of \emph{quantum quenches}~\cite{cc-06}, the generic growth and late-time saturation of entanglement measures correspond to the emergence of a local stationary mixed state~\cite{rigol2008thermalization,calabrese2016quantum,essler2016quench,vidmar2016generalized}.

The connections between entanglement and thermodynamics can be studied very explicitly in one-dimensional integrable systems~\cite{korepin1993}, which are very useful toy models with an extensive number of local conservation laws. As a defining feature, their Hamiltonian can be diagonalized analytically, with the corresponding energy spectrum admitting an intuitive description in terms of stable quasiparticle excitations. It has been known for a long time that their macrostates are fully described by the corresponding distribution function of quasiparticle quasimomenta~\cite{takahashi1999thermodynamics}, in analogy to non-interacting quantum gases. For example, at thermal equilibrium the system entropy is given by the \emph{Yang-Yang entropy} \cite{yy-69}, a simple functional of such distribution functions~\cite{takahashi1999thermodynamics}. As an important milestone, it has been recently realized that a similar description also holds out of equilibrium. In particular, in an appropriate scaling limit, the entanglement dynamics after a quench can be captured by a semiclassical formula featuring the Yang-Yang entropy corresponding to the post-quench stationary state~\cite{fagotti2008evolution,alba2017entanglement,alba2018entanglement,c-20} combined with the quasiparticle picture of entanglement spreading \cite{cc-05}. 
While not yet proven in general, this formula has been numerically tested in a number of cases~\cite{alba2017entanglement,alba2018entanglement,mestyan2018renyi,modak2019correlation, lagnese2021entanglement,ac-17b}, and derived analytically in non-interacting systems~\cite{fagotti2008evolution}, as well as in certain integrable quantum cellular automata~\cite{klobas2021thermalization,klobas2021entanglement,klobas2021exact_II}.

More recently, the interplay between entanglement entropies and symmetries has attracted increasing theoretical~\cite{laflorencie2014spin,goldstein2018symmetry,xavier2018,neven2021symmetry} and experimental~\cite{lukin2019probing,vitale2022symmetry} interest, with a lot of work being devoted to the study of \emph{symmetry-resolved entanglement entropies} (SRRE). Most of the existing literature has so far focused on ground-state physics in various theoretical contexts, such as CFTs~\cite{goldstein2018symmetry, xavier2018,cornfeld2018imbalance,murciano2021symmetry,chen2022charged,chen2021symmetry,capizzi2021symmetry,hung2021entanglement,calabrese2021symmetry,capizzi2020symmetry,bonsignori2020boundary,estienne2021finite,ghasemi2022universal}, free~\cite{murciano2020entanglement,Horvath2021symmetry} and interacting integrable quantum field theories~\cite{horvath2020symmetry,capizzi2021entanglement,horvath2021branch}, holographic settings~\cite{belin2013holographic,caputa2016charged,zhao2021symmetry,weisenberger2021symmetry}, spin chains~\cite{j-22,bonsignori2019symmetry,fraenkel2020symmetry,feldman2019dynamics,murciano_digiulio_2020symmetry,murciano2020symmetry,calabrese2020full,ares2022symmetry,wiseman2003entanglement,barghathi2018renyi,barghathi2019operationally,barghathi2020theory,murciano2020symmetry,tan2020particle,ma2022symmetric}, disordered systems~\cite{turkeshi2020entanglement,kiefer2020bounds,kiefer2021slow,kiefer2020evidence} and non-trivial topological
phases~\cite{cornfeld2019entanglement,monkman2020operational,azses2020symmetry,ore-21,ahn-20,ads-21}. On the other hand, only a few results have been derived for thermal states~\cite{ghasemi2022universal} and out-of-equilibrium situations~\cite{lukin2019probing,parez2021exact,parez2021quasiparticle,fraenkel2021entanglement,parez2022dynamics}. 

In the context of quantum quenches, Refs.~\cite{parez2021exact,parez2021quasiparticle} derived analytic results for the dynamics of the SREE following a quench in non-interacting fermionic systems, and showed how quantitative predictions could be obtained generalizing the quasiparticle picture developed for the standard entanglement dynamics~\cite{fagotti2008evolution}. While extending the exact microscopic calculations to interacting integrable models appears to be out of reach, one could wonder whether it is possible to generalize the semiclassical formula beyond the non-interacting regime, thus obtaining a conjecture for the SREE analogous to that of Ref.~\cite{alba2017entanglement} for the standard entanglement entropy.

A first step in this direction is to compute the SREE for the post-quench stationary state. Assuming the validity of the generalized Gibbs ensemble (GGE)~\cite{vidmar2016generalized,essler2016quench}, this amounts to understand how the SREE of a generic macrostate can be related to the corresponding distribution function of quasiparticle quasimomenta. Indeed, in the case of the standard entanglement entropy, the identification of the latter with the Yang-Yang entropy played a crucial role for the heuristic derivation of the formula put forward in Ref.~\cite{alba2017entanglement}.

Motivated by these discussions, in this work we develop a general approach to compute the symmetry-resolved R\'enyi and von Neumann entanglement entropies of thermodynamic macrostates in interacting integrable systems. For concreteness, we will focus on the prototypical example of the XXZ Heisenberg spin chain, although our method can be clearly extended to any Bethe ansatz integrable model. Our approach is based on a combination of the thermodynamic Bethe ansatz~\cite{takahashi1999thermodynamics} and the G\"artner-Ellis theorem from large deviation theory~\cite{Touchette2009Large}. As our main result, we will show that the von Neumann SREE coincides with the Yang-Yang entropy of an effective macrostate determined by the charge sector.

The rest of this work is organized as follows. In Sec.~\ref{sec:SREE} we define the symmetry-resolved R\'enyi and von Neumann entanglement entropies. In Sec.~\ref{sec:model} we introduce the XXZ Heisenberg spin chain and review its description in terms of the thermodynamic Bethe ansatz. Next, in Sec.~\ref{sec:SREE_large_dev} we discuss how the SREE can be obtained in terms of certain \emph{rate functions} within large deviation theory and show that the thermodynamic Bethe ansatz can be conveniently used to explicitly compute them. Finally, in Sec.~\ref{sec:applications}, we present two applications of this formalism, presenting results for the SREE of thermal states and GGEs corresponding to a quench from tilted Néel states. Our conclusions are consigned to Sec.~\ref{sec:conclusions}.

\section{Symmetry resolved entanglement entropies}
\label{sec:SREE}

Let us begin by reviewing the definitions of symmetry-resolved R\'enyi and von Neumann entanglement entropies. We consider a system with a $U(1)$-charge operator $\hat{Q}$, in a given state described by the density matrix $\varrho$. Denoting by $A$, $\bar{A}$ a spatial bipartition, we assume $\hat{Q}=\hat{Q}_{A}\oplus \hat{Q}_{\bar{A}}$. If $[\varrho,\hat{Q}]=0$, then  $[\varrho_{A},\hat{Q}_A]=0$, where $\varrho_{A}={\rm tr}_{\bar{A}}[\varrho]$ is the reduced density matrix of the subsystem $A$. Accordingly, $\varrho_{A}$ is block-diagonal, and we can write
\be
\varrho_{A}=\oplus_{Q} p_{A}(Q) \varrho_{A}(Q)\,,
\ee
where $p_{A}(Q) ={\rm tr}\left[\Pi_{Q}\rho_A\right]$, and $\Pi_Q$ is the projector onto the sector of charge $Q$ in the subsystem $A$.
Introducing the von Neumann entanglement entropy
\be
S_{1}(\varrho_A)=-{\rm tr}\left[\varrho_A\ln\varrho_A\right]\,,
\ee
a simple calculation shows
\be
S_{1}(\varrho_A)=\sum_{Q} p_A(Q) S_{1}(Q)-\sum_{Q} p_A(Q) \ln p_A(Q)\,,
\ee
where we introduced the symmetry resolved entanglement entropy
\be
S_{1}(Q) \equiv-\operatorname{tr}\left[ \varrho_{A}(Q) \ln \varrho_{A}(Q)\right]\,.
\ee
Similarly, the symmetry resolved R\'enyi entropies can be defined as
\be
S_{\alpha}(Q) \equiv \frac{1}{1-\alpha} \ln {\rm tr} [\varrho_{A}^{\alpha}(Q)]\,.
\ee

If the reduced density matrix $\varrho_A$ is known exactly, a simple strategy to compute the SREE was put forward  in Ref.~\cite{goldstein2018symmetry,xavier2018}. Namely, one first computes the charged moments
\be\label{eq:charged_moments}
Z_{\alpha}(\theta) \equiv \operatorname{tr}\left[\varrho_{A}^{\alpha} \mathrm{e}^{\mathrm{i} \theta \hat{Q}_{A}}\right]\,,
\ee
and their Fourier transform
\be\label{eq:fourier}
\mathcal{Z}_{\alpha}(Q)=\int_{-\pi}^{\pi} \frac{d \theta}{2 \pi} \mathrm{e}^{-\mathrm{i} Q \theta} Z_{\alpha}(\theta) \equiv \operatorname{tr}\left[\Pi_{Q} \varrho_{A}^{\alpha}\right]\,.
\ee
The symmetry-resolved R\'enyi and von Neumann entanglement entropies are then obtained as
\be
S_{\alpha}(Q)=\frac{1}{1-\alpha} \ln \left[\frac{\mathcal{Z}_{\alpha}(Q)}{\mathcal{Z}_{1}(Q)^{\alpha}}\right], \quad S_{1}(Q) = \lim_{\alpha\to 1} S_\alpha(Q)=-\partial_{\alpha}\left[\frac{\mathcal{Z}_{\alpha}(Q)}{\mathcal{Z}_{1}(Q)^{\alpha}}\right]_{\alpha=1}\,,
\ee
while 
\be\label{eq:pQ_moments}
p_A(Q)=\mathcal{Z}_{1}(Q)\,.
\ee

\section{The model and the thermodynamic Bethe ansatz}
\label{sec:model}

As mentioned, we will focus on the XXZ Heisenberg spin chain
\be
\label{eq:hamiltonian}
H=\frac{1}{4}\sum_{j=1}^L\left[\sigma_j^x\sigma_{j+1}^x+\sigma_j^y\sigma_{j+1}^y+\Delta\sigma_j^z\sigma_{j+1}^z\right]\,,
\ee
where $\sigma^\alpha$ are the Pauli matrices, while $\Delta$ is the anisotropy parameter. For concreteness, we will restrict to the gapped regime of the model, $\Delta>1$, and set $\Delta=\cosh(\eta)$ with $\eta\in\mathbb{R}$ (the generalization to the case $\Delta<1$ is straightforward). For the Hamiltonian~\eqref{eq:hamiltonian} the $U(1)$-symmetry charge is given by the magnetization
\be\label{eq:magnetization}
S^{z}=\frac{1}{2}\sum_{j=1}^L\sigma_j^z\,.
\ee
We will take $L$ even, so that the eigenvalues of $S^{z}$ are integer numbers. 

Due to its underlying integrability, the model features an infinite number of pairwise commuting local and quasilocal conserved operators (charges), see Ref.~\cite{ilievski2016quasilocal} for a comprehensive review. For $\Delta>1$, they are all even under spin inversion, except for the magnetization~\eqref{eq:magnetization}. Following~\cite{ilievski2016quasilocal}, we denote them by $Q_{s}^{(j)}$, where $s = 1/2, 1, 3/2, 2,\ldots $ and $j = 1, 2,\ldots $. With this convention, $Q^{(j)}_{1/2}$ are strictly local, and $Q^{(2)}_{1/2}\propto H$~\cite{ilievski2016quasilocal}. 

The Hamiltonian~\eqref{eq:hamiltonian} can be diagonalized exactly using the Bethe ansatz. Each eigenstate is associated with a set of quasiparticle quasimomenta, or \emph{rapidities}, $\{\lambda_j\}_{j}$, satisfying a set of quantization conditions known as Bethe equations~\cite{korepin1993}. In the thermodynamic limit, the rapidities arrange themselves into patterns in the complex plane and their collective behavior can be described in terms of rapidity distribution functions~\cite{takahashi1999thermodynamics}. In particular, in the regime $\Delta>1$, a given state is characterized by the set $\boldsymbol{\rho}\equiv\{\rho_n(\lambda)\}_{n=1}^\infty$, where $\lambda\in[-\pi/2,\pi/2]$ while the index $n$ labels different bound-states of quasiparticles. Together with $\rho_n(\lambda)$, one also needs to introduce the distribution functions $\rho^h_n(\lambda)$ for the unoccupied modes, or ``holes'', satisfying~\cite{takahashi1999thermodynamics}
\be
\rho_{m}(\lambda)+\rho_{m}^{h}(\lambda)=a_{m}(\lambda)-\sum_{n=1}^{\infty}\left[a_{m n} * \rho_{n}\right](\lambda)\,,
\label{eq:Bethe_equations}
\ee
where we defined the convolution
\be
(f * g)(\lambda)=\int_{-\pi / 2}^{\pi / 2} d \mu f(\lambda-\mu) g(\mu)\,,
\ee
and 
\bea
a_{m n}(\lambda)&=&\left(1-\delta_{m n}\right) a_{|m-n|}(\lambda)+2 a_{|m-n|+2}(\lambda)+\ldots+2 a_{m+n-2}(\lambda)+a_{m+n}(\lambda)\,,\\
a_{n}(\lambda)&=&\frac{1}{\pi} \frac{\sinh (n \eta)}{\cosh (n \eta)-\cos (2 \lambda)}\,.
\eea
From the knowledge of the densities $\{\rho_n(\lambda)\}$, several thermodynamic quantities can be computed immediately. For instance, the correspondent magnetization and energy densities are
\begin{align}
\mathcal{S}^{z}[\boldsymbol{\rho}]&=\frac{1}{2}-\sum_{n=1}^\infty n\int_{-\pi / 2}^{\pi / 2}  \mathrm{d} \lambda~\rho_{n}(\lambda)\,,\label{eq:s_z}\\
\mathcal{H}[\boldsymbol{\rho}]&=\sum_{n=1}^\infty \int_{-\pi / 2}^{\pi / 2}  \mathrm{d} \lambda~\rho_{n}(\lambda)\varepsilon_n(\lambda)\,,
\end{align}
where
\be
\varepsilon_n(\lambda)=-\pi\sinh(\eta)a_{n}(\lambda)\,.
\ee
Similarly, the density of any quasilocal charge can be written as
\be
\mathcal{Q}^{(j)}_s[\boldsymbol{\rho}]=\sum_{n=1}^\infty\int_{-\pi / 2}^{\pi / 2}  {\rm d}\lambda~\rho_n(\lambda)q_{s,n}^{(j)}(\lambda)\,,
\ee
where $q_{s,n}^{(j)}(\lambda)$ are known functions~\cite{ilievski2016quasilocal}.

The solution to the set of equations~\eqref{eq:Bethe_equations} is not unique. This corresponds to the fact that the model admits infinitely many equilibrium stationary states, which can be represented in the grand canonical form
\be\label{eq:GGE_rho}
\varrho_{\mathrm{GGE}} \equiv \frac{1}{Z_{\mathrm{GGE}}} \exp \left(hS^z-\sum_{s, j} \beta_{s}^{(j)} Q_{s}^{(j)}\right)\,,
\ee
where 
\be
Z_{\mathrm{GGE}} \equiv \operatorname{tr}\left[ \exp \left(hS^z-\sum_{s, j} \beta_{s}^{(j)} Q_{s}^{(j)}\right)\right]\,
\ee
is a normalization. The density matrices~\eqref{eq:GGE_rho} describe the set of generalized Gibbs ensembles, which depend on the Lagrange multipliers $h$, $\beta_s^{(j)}$. As explained in Ref.~\cite{mossel2012journal} (see also Refs.~\cite{alba2017quench,mestyan2018renyi}) the corresponding rapidity distribution functions $\rho_n(\lambda)$ can be found following the standard prescriptions of the thermodynamic Bethe ansatz~\cite{takahashi1999thermodynamics} (TBA). In particular, the function $\rho_n(\lambda)$ is given by the saddle-point condition
\be\label{eq:saddle-point}
\frac{\delta}{\delta \rho_n}\mathcal{F}\left( h, \beta^{(j)}_{s} , \boldsymbol{\rho} \right)=0\,,
\ee
where we have introduced the generalized free energy density
\be\label{eq:generalized_free_energy}
\mathcal{F}\left( h, \beta^{(j)}_{s} , \boldsymbol{\rho} \right)=-\mathcal{E}[h, \beta^{(j)}_{s}, \boldsymbol{\rho} ]+\mathcal{S}_{\rm YY}[ \boldsymbol{\rho}]\,.
\ee 
Here we defined
\begin{align}
\mathcal{E}[h,\beta_s^{(j)}, \boldsymbol{\rho}] &\equiv -h \mathcal{S}^z[\boldsymbol{\rho}]+\sum_{s, j} \beta_{s}^{(j)} \mathcal{Q}_{s}^{(j)}[\boldsymbol{\rho}]
\nonumber\\
&= -h\left(\frac{1}{2}-\sum_{n} n\int \mathrm{d} \lambda~ \rho_{n}(\lambda)\right) +\sum_{s, j} \beta_{s}^{(j)} \sum_{n} \int \mathrm{d} \lambda~ \rho_{n}(\lambda) q_{s, n}^{(j)}(\lambda)\,,
\end{align}
while
\be
\mathcal{S}_{\mathrm{YY}}[\boldsymbol{\rho}]= \sum_{n=1}^{\infty} \int \mathrm{d} \lambda~\left[\rho_{t, n}(\lambda) \ln \rho_{t, n}(\lambda)-\rho_{n}(\lambda) \ln \rho_{n}(\lambda)-\rho_{h, n}(\lambda) \ln \rho_{h, n}(\lambda)\right]\,
\ee
is the Yang-Yang entropy density. Introducing the auxiliary functions
\be
\eta_n(\lambda)=\frac{\rho^h_n(\lambda)}{\rho_n(\lambda)}\,,
\ee
and after standard manipulations, the saddle-point condition can be rewritten as
\be\label{eq:eta_functions}
\ln \eta_{n}(\lambda)= nh+ g_{n}(\lambda)+\sum_{m=1}^{\infty} a_{n m} \ast \ln \left[1+1 / \eta_{m}\right](\lambda)\,,
\ee
with
\be\label{eq:gn_functions}
g_{n}(\lambda) \equiv \sum_{s, j} \beta_{s}^{(j)} q_{s, n}^{(j)}(\lambda)\,.
\ee
The set of equations~\eqref{eq:eta_functions} can be solved numerically by simple iterative methods. Once the functions $\eta_n(\lambda)$ are known, one can use $\rho_n^h(\lambda)=\eta_n(\lambda)\rho_n(\lambda)$ and solve~\eqref{eq:Bethe_equations} for the functions $\rho_n(\lambda)$, finally obtaining the rapidity distribution functions corresponding to the state~\eqref{eq:GGE_rho}. 

\section{SREE and large-deviation theory}
\label{sec:SREE_large_dev}

In the rest of this work, we aim to compute the SREE assuming that the subsystem reduced density matrix has the form~\eqref{eq:GGE_rho}. Note that, in principle, one should take into account boundary effects, because the definition~\eqref{eq:GGE_rho} implicitly assumes an infinite system-size limit. However, we will only be interested in the leading extensive behavior of the SREE. Therefore, we will neglect any boundary effect, because they are expected to give sub-leading contributions.

In principle, the SREE could be computed using the strategy outlined in Sec.~\ref{sec:SREE}, cf. Eqs.~\eqref{eq:charged_moments}--\eqref{eq:pQ_moments}. However, for thermodynamic states, the charged moment are expected to be exponentially small in the subsystem size, making the computation of the Fourier transform in~\eqref{eq:fourier} challenging from the numerical point of view. Here we provide an alternative approach, which is tailored to capture the leading behavior of the SREE and which turns out to be particularly convenient to be implemented within the TBA formalism. 

Our method is based on an application of the G\"artner-Ellis theorem from large deviation theory~\cite{Touchette2009Large} and it is valid beyond integrable models. We introduce the general idea in Sec.~\ref{sec:general_idea}, while in Sec.~\ref{sec:application_TBA} we discuss its implementation within the TBA formalism. 

\subsection{Extensive SREE from the G\"artner-Ellis theorem}
\label{sec:general_idea}
Let $\varrho_A$ be the reduced density matrix associated with a subsystem $A$, containing $\ell$ spins.  Since $[\varrho,\Pi_Q]=0$ and $\Pi_Q^\alpha=\Pi_Q$ for any power $\alpha$, we can write
\be\label{eq:p-s_decomposition}
{\rm tr} [\varrho^{\alpha}(Q)]=\frac{{\rm tr} [\Pi_Q\varrho^{\alpha}]}{{\rm tr}[ \Pi_Q \varrho]^{\alpha}}=\frac{{\rm tr} [\Pi_Q\varrho^{\alpha}]}{{\rm tr}[\varrho^\alpha]}\frac{{\rm tr}[\varrho^\alpha]}{(p_1(Q))^\alpha}=p_\alpha(Q) \frac{{\rm tr}[\varrho^\alpha]}{(p_1(Q))^\alpha}\,,
\ee
where we have introduced
\be
p_\alpha(Q)=\frac{{\rm tr} [\Pi_Q\varrho^{\alpha}]}{{\rm tr}[\varrho^\alpha]}\,.
\ee
Clearly, $p_\alpha(Q)\geq 0$ and, using standard considerations, it is easy to see that $\sum_{Q} p_\alpha(Q)=1$, i.e. $p_\alpha(Q)$ can be interpreted as a probability distribution. Setting $q=Q/\ell$, we expect on physical grounds that $p_\alpha(Q)$ should follow a large deviation principle in the large-$\ell$ limit, that is,
\be
p_\alpha(Q)\sim e^{-I_{\alpha}(q)\ell}\,,
\ee
where $I_\alpha(q)$ is referred to as the \emph{rate function}. In this setting, it is well-known that $I_\alpha(q)$ can be computed by means of the G\"artner–Ellis theorem~\cite{Touchette2009Large}. To this end, we define the generating function
\be
G_\alpha(w)=\frac{{\rm tr}\left[ \varrho^{\alpha} e^{w \hat{Q}}\right]}{{\rm tr}[\varrho^\alpha]}\sim e^{\ell f_\alpha(w)}\,,\qquad w\in \mathbb{R}\,.
\ee
The G\"artner–Ellis theorem states that we can compute $I_\alpha(q)$ as the Legendre–Fenchel transform of  $f_\alpha(w)$. In the case where the functions involved are concave and regular, the latter reduces to the Legendre transform, and we can write
\be\label{eq:rate_function_alpha}
I_\alpha(q)=w_{\alpha, q} q-f_\alpha\left(w_{\alpha, q}\right)\,,
\ee
where $w_{\alpha, q}$ is determined by the condition
\be
\left.\frac{d}{d w}(f_\alpha(w)-w q)\right|_{w=w_{\alpha, q}}=0\,.
\label{eq:sqdefinition_alpha}
\ee

Now, defining the density of SREE entropy
\be
s_\alpha(q)=\lim_{\ell\to\infty} \frac{S_\alpha(q\ell)}{\ell} 
= \frac{1}{1-\alpha}\lim_{\ell\to\infty} \frac{\ln {\rm tr} [\varrho^\alpha(q \ell)]}{\ell}
\ee
and using~\eqref{eq:p-s_decomposition}, we obtain
\be\label{eq:final_result_alpha}
s_\alpha(q)=s_\alpha+\frac{1}{1-\alpha}\left[-I_{\alpha}(q)+\alpha I_{1}(q)\right]\,,
\ee
where 
\be 
s_{\alpha} =  \lim_{\ell\to\infty} \frac{1}{1-\alpha}\frac{\ln {\rm tr} [\varrho^\alpha]}{\ell} 
\ee
is the density of Rényi entropy. We may now take the limit $\alpha\to1$ to recover the von Neumann SREE, yielding
\be\label{eq:final_result_vN}
s_1(q)=s_1+I_1(q)+\frac{{\rm d} I_\alpha(q)}{{\rm d}\alpha}\Big|_{\alpha=1}\,.
\ee

\subsection{Explicit computations via TBA}
\label{sec:application_TBA}

In this section we show how the program presented above can be followed within the thermodynamic Bethe ansatz formalism, arriving at an explicit exact formula for the SREE. At the technical level, similar calculations appeared, albeit in a completely different context, in Ref.~\cite{PePi19}, where, following Refs.~\cite{silva2008statistics,gambassi2012large}, the G\"artner-Ellis theorem was applied to the study of the work statistics after a quench in the integrable Lieb-Liniger model.

We start by computing
\be 
f_\alpha(w) = \lim_{\ell \to \infty}\frac{1}{\ell} \ln \frac{{\rm tr} (\varrho^{\alpha} e^{w \hat{Q}})}{{\rm tr}[\varrho^{\alpha}]}\,.
\ee
Let us recall that $\varrho$ denotes in principle the reduced density matrix in a subsystem of size $\ell$, within a system of total size $L$. However, since we are interested in the extensive part of the entanglement entropies, it can be argued that the calculations amount to taking first $\ell=L$, and then the limit $L\to \infty$. Hence, $\varrho$ can be taken to be of the same form as the full system's density matrix (the same assumption has been used in Refs.~\cite{alba2017quench,mestyan2018renyi} for the computation of the post-quench R\'enyi entropies).

Assuming that $\varrho$ is of the form~\eqref{eq:GGE_rho}, the density matrix $\varrho^{\alpha}$ is of the same form, but with Lagrange multipliers $\alpha h$,  $\alpha\beta_s^{(j)}$. Following the standard prescription~\cite{takahashi1999thermodynamics,mossel2012journal}, we can therefore cast the trace into a functional integral over rapidity distribution functions, obtaining 
\be 
f_\alpha(w) = \mathcal{F}\left( w +\alpha h , \alpha \beta^{(j)}_{s}  , \boldsymbol{\rho}[w +\alpha h ,\alpha \beta^{(j)}_{s}] \right)
- \mathcal{F}\left(\alpha h, \alpha \beta^{(j)}_{s} , \boldsymbol{\rho}[\alpha h ,\alpha \beta^{(j)}_{s}]  \right)\,,
\ee 
where $\mathcal{F}(h,\beta^{(j)}_s,\boldsymbol{\rho})$ is defined in~\eqref{eq:generalized_free_energy}, while we denoted by $\boldsymbol{\rho}[h,\beta^{(j)}_s]$ the set of saddle-point rapidity distribution functions associated with the GGE~\eqref{eq:GGE_rho} with Lagrange multipliers $h$, $\beta^{(j)}_s$. Next, by using the definition of $\boldsymbol{\rho}[w +\alpha h ,\alpha \beta^{(j)}_{s}]$ in terms of a saddle-point condition~\eqref{eq:saddle-point}, it is straightforward to see that~\eqref{eq:sqdefinition_alpha} can be rewritten as
\be 
\mathcal{S}^z[ \boldsymbol{\rho}[w_{\alpha, q} +\alpha h ,\alpha \beta^{(j)}_{s}]]=q\,,
\label{eq:density_condition}
\ee
where $\mathcal{S}^{s}[\boldsymbol{\rho}]$ was defined in~\eqref{eq:s_z}. Finally, using~\eqref{eq:rate_function_alpha} and \eqref{eq:final_result_alpha}, we arrive at
\bea\label{eq:final_renyi}
s_\alpha(q) &=&     \frac{1}{1-\alpha} \left( -\mathcal{E}[\alpha h, \alpha\beta^{(j)}_s, \boldsymbol{\rho}[w_{\alpha, q} +\alpha h, \alpha \beta^{(j)}_s]]+\mathcal{S}_{YY}[\boldsymbol{\rho}[w_{\alpha, q} +\alpha h, \alpha \beta^{(j)}_s]]+\right. \nonumber \\
&+&
\left.
\alpha \mathcal{E}[ h, \beta^{(j)}_s, \boldsymbol{\rho}[w_{1, q} + h,  \beta^{(j)}_s]]- \alpha \mathcal{S}_{YY}[ \boldsymbol{\rho}[w_{1, q} + h,  \beta^{(j)}_s]]
\right)\,.
\eea
This formula gives us the density of the R\'enyi SREE. Note that $w_{\alpha,q}$, defined by ~\eqref{eq:density_condition}, can be interpreted as an additional Lagrange multiplier in the GGE, introduced in order to make the magnetization equal to $q$. 

Let us evaluate~\eqref{eq:final_renyi} in the limit $\alpha\to 1$. We expand
\be
 -\mathcal{E}[\alpha h, \alpha\beta^{(j)}_s, \boldsymbol{\rho}[w_{\alpha, q} +\alpha h, \alpha \beta^{(j)}_s]]+\mathcal{S}_{YY}[\boldsymbol{\rho}[w_{\alpha, q} +\alpha h, \alpha \beta^{(j)}_s]]=t_1+(\alpha-1)t_2+o(1-\alpha)\,.
\ee
Clearly
\be
t_1= -\mathcal{E}[ h, \beta^{(j)}_s, \boldsymbol{\rho}[w_{1, q} + h,  \beta^{(j)}_s]]+ \mathcal{S}_{YY}[ \boldsymbol{\rho}[w_{1, q} + h,  \beta^{(j)}_s]]\,.
\ee
Next, writing
\be
 -\mathcal{E}[\alpha h, \alpha\beta^{(j)}_s, \boldsymbol{\rho}[w_{\alpha, q} +\alpha h, \alpha \beta^{(j)}_s]]=w_{\alpha,q} q-\mathcal{E}[w_{\alpha, q}+\alpha h, \alpha\beta^{(j)}_s, \boldsymbol{\rho}[w_{\alpha, q} +\alpha h, \alpha \beta^{(j)}_s]]\,,
\ee 
recalling the definition of $\boldsymbol{\rho}[w_{1, q} + h,  \beta^{(j)}_s]$ in terms the saddle-point condition, and using 
\be
\frac{d}{d \alpha}=\frac{\partial}{\partial \alpha}  + \frac{\delta}{\delta \boldsymbol{\rho}}  \frac{\partial \boldsymbol{\rho}}{\partial \alpha}\,,
\ee
we get
\be
t_2=-\mathcal{E}[ h, \beta^{(j)}_s, \boldsymbol{\rho}[w_{1, q} + h,  \beta^{(j)}_s]]\,.
\ee
Putting everything together, we finally obtain
\be\label{eq:main_result}
s_{1}(q)=\mathcal{S}_{YY}[\boldsymbol{\rho}[w_{1, q} + h,  \beta^{(j)}_s]]]\,.
\ee

Eq.~\eqref{eq:main_result} is the main result of this section. It states that the value of the density of the von Neumann SREE can be obtained as the Yang-Yang entropy density associated with the macrostate $\boldsymbol{\rho}[w_{1, q} + h,  \beta^{(j)}_s]$, whose magnetization must be equal to $q$ (because of the condition~\eqref{eq:density_condition} defining $w_{1,q}$). Note that, in order to evaluate numerically Eqs.~\eqref{eq:final_renyi} and \eqref{eq:main_result}, one needs to find the value of $w_{\alpha,q}$ for which~\eqref{eq:density_condition} is satisfied, which can be achieved by standard numerical procedures. 

\section{Applications}
\label{sec:applications}

In this section we apply the formalism developed in the previous sections and provide results for the SREE in thermal states and in the GGE following a quantum quench from tilted Néel states.

\subsection{Thermal states}

\begin{figure}
	\begin{tabular}{lll}
		\hspace{-0.15cm}\includegraphics[width=0.33\textwidth]{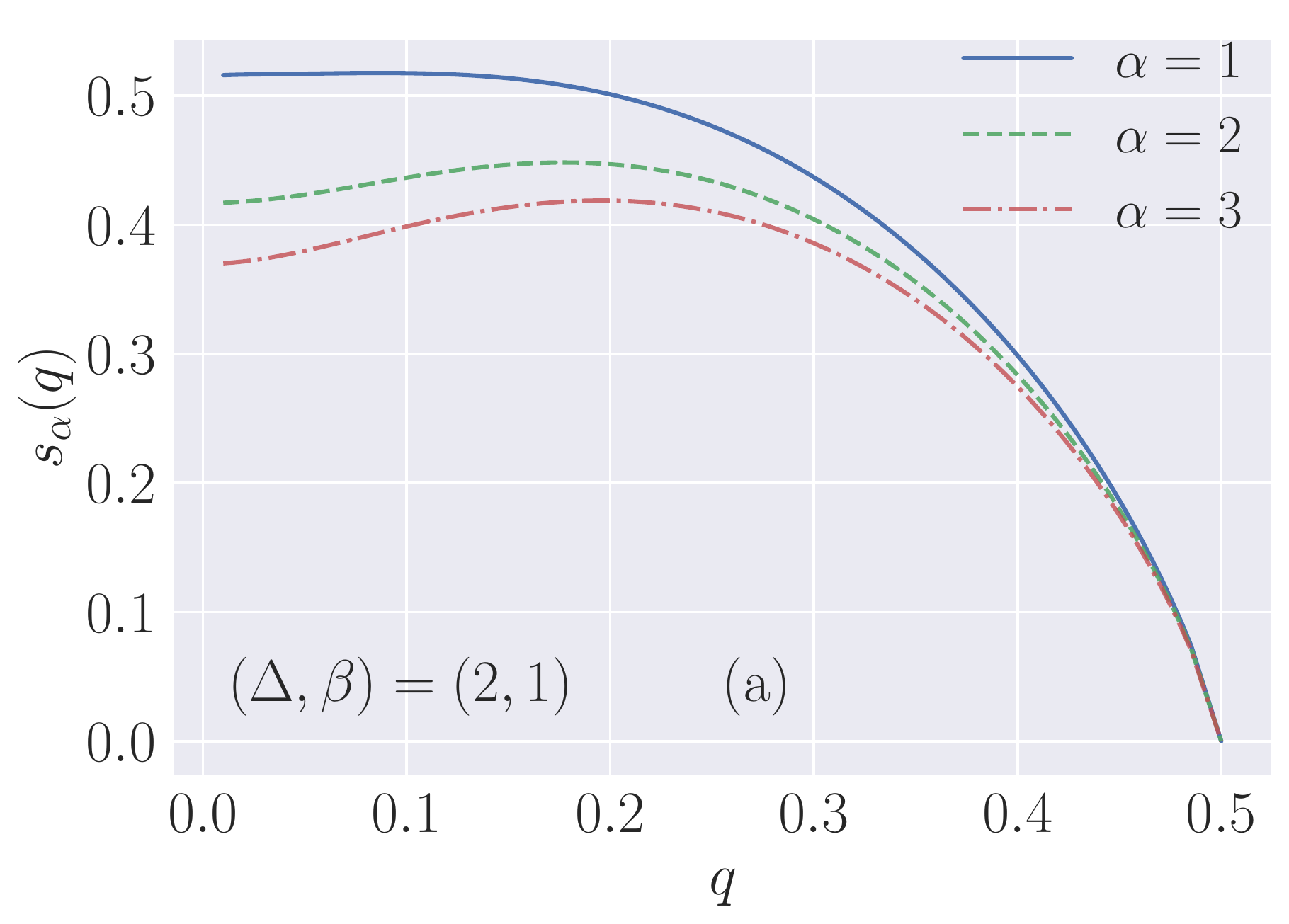} & \hspace{-0.15cm}\includegraphics[width=0.33\textwidth]{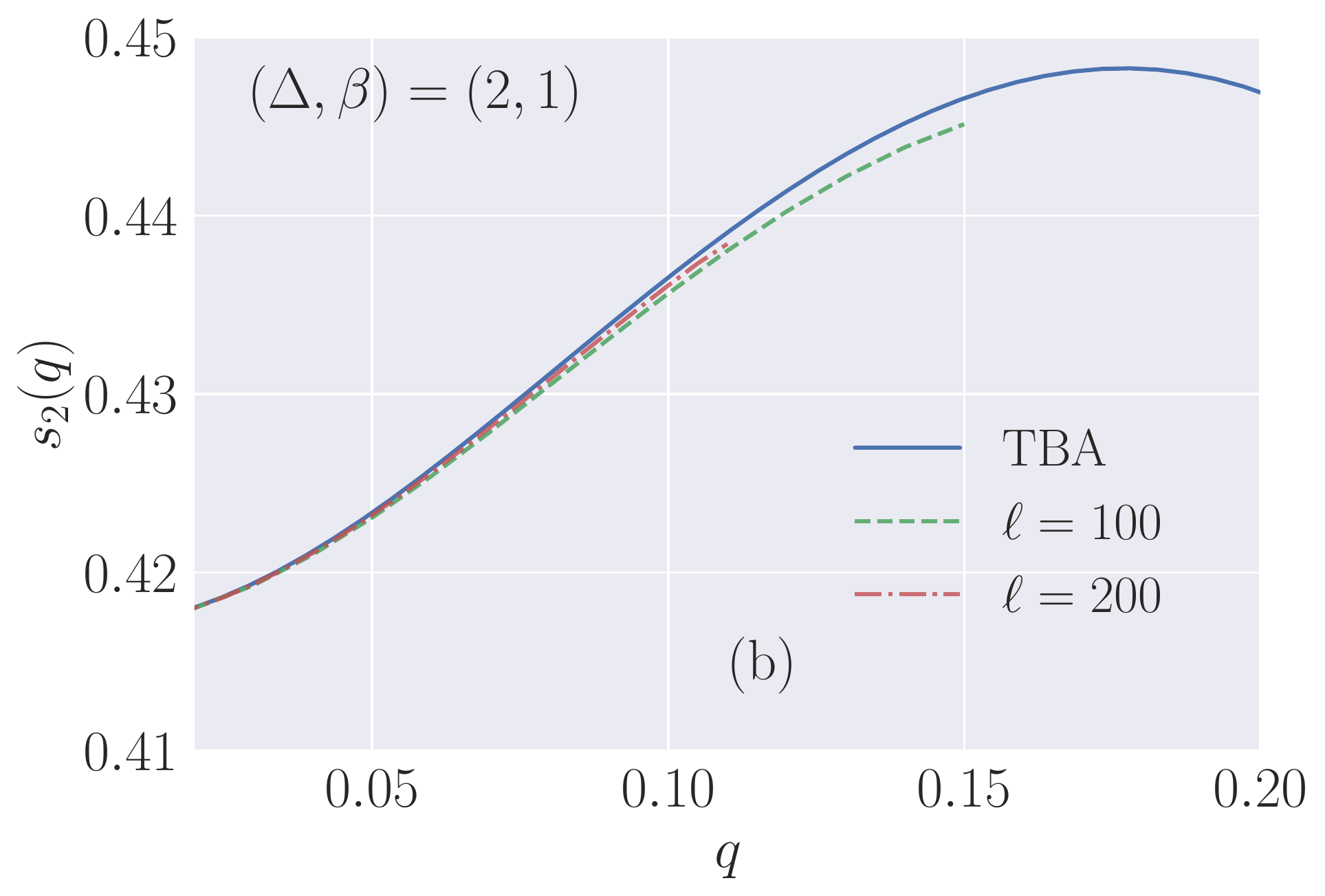}& \hspace{-0.15cm}\includegraphics[width=0.33\textwidth]{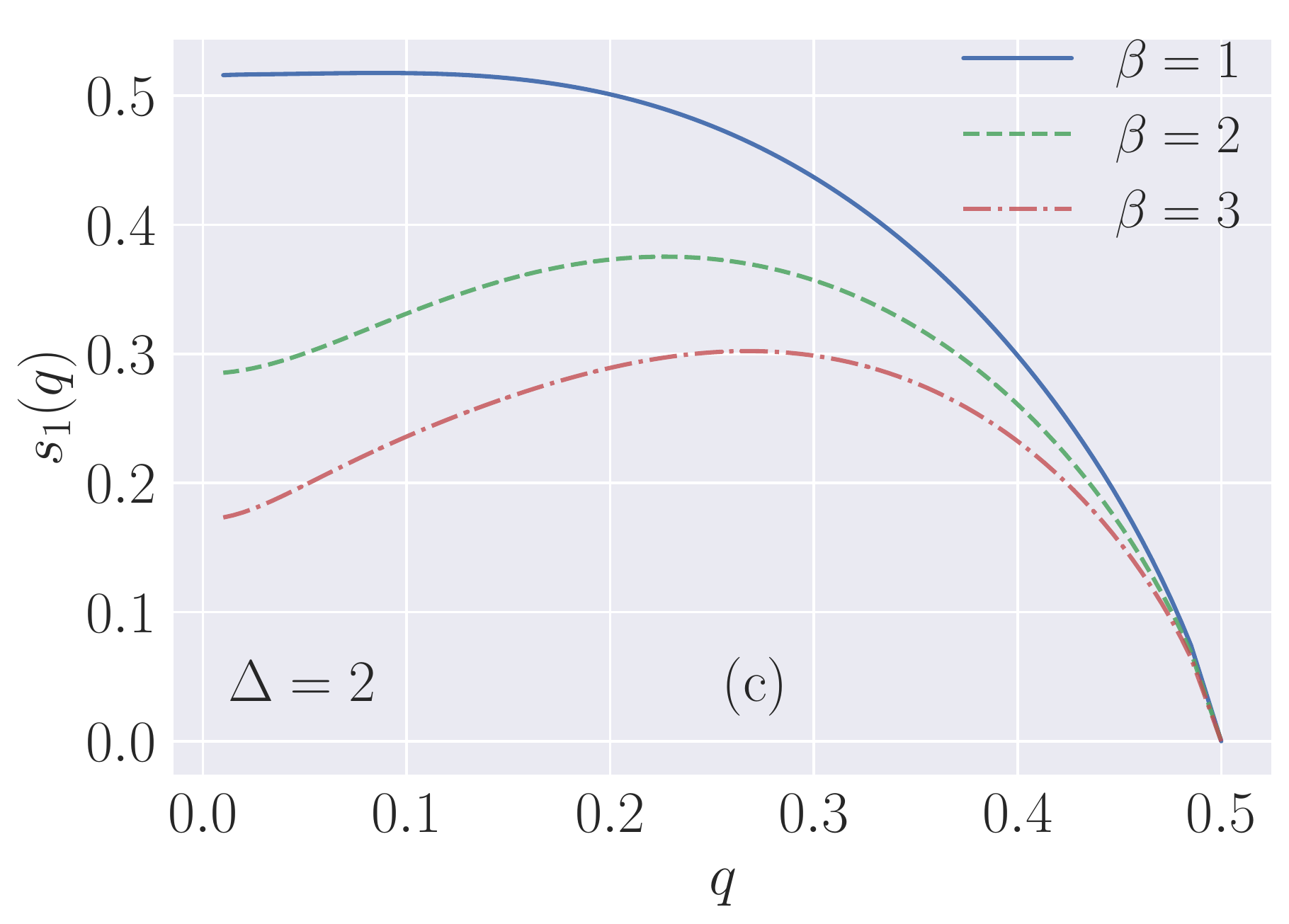}\\
	\end{tabular}
	\caption{Left: Density of symmetry resolved R\'enyi entanglement entropy, corresponding to the thermal state~\eqref{eq:thermal_state}, with $\beta=1$ and $\Delta=2$. Center: comparison between analytical TBA results and numerical data obtained by iTEBD calculations for the R\'enyi-$2$ SREE. As $\ell$ increases, we were only able to obtain reliable iTEBD data for smaller values of $q$ (cf. the main text). Right: Density of symmetry resolved von Neumann entanglement entropy. In all plots, we only show results for $q>0$, the curves being manifestly symmetric for $q<0$.}
	\label{fig:sree_thermal}
\end{figure}

Let us first consider the case in which the reduced density matrix is described by a thermal state
\be\label{eq:thermal_state}
\varrho=\frac{e^{-\beta H}}{\mathcal{Z}}\,.
\ee
The rapidity distribution functions are determined by the system of equations~\eqref{eq:eta_functions} where now~\cite{takahashi1999thermodynamics}
\be
g_n(\lambda)=-\beta\pi\sinh\eta a_n(\lambda)\,.
\ee
We have solved numerically the TBA equations and evaluated Eqs.~\eqref{eq:final_renyi} and \eqref{eq:main_result}. Examples of our data are reported in Fig.~\ref{fig:sree_thermal}. We observe a generic non-monotonic behavior for both the R\'enyi and von Neumann SREE, which is more pronounced for smaller values of the temperature (i.e. larger values of $\beta$) and larger R\'enyi index. 

Obtaining an independent numerical confirmation of our results appears to be challenging. This is because computations based on exact-diagonalization techniques are restricted to relatively small system sizes, which are plagued by large finite-size effects. To make things worse, the dimension of a given charge sector is smaller than that of the whole Hilbert space. In order to get around this issue, we have performed numerical computations based on the iTEBD algorithm~\cite{vidal2007classical}, which allow us to reach large sub-system sizes. More precisely, we have computed numerically the reduced density matrix over $\ell$ sites, starting from an infinite system prepared in a thermal state at a given temperature $\beta^{-1}$, and extracted the SREE using the strategy outlined in Sec.~\ref{sec:SREE}, cf. Eqs.~\eqref{eq:charged_moments}--\eqref{eq:pQ_moments}.

As a general numerical limitation of this approach, the charged moments are exponentially small in $\ell$ and it is therefore difficult to obtain data with good relative precision for large values of $\alpha$ in~\eqref{eq:charged_moments}. This results in significant numerical inaccuracies for the density of SREE for large values of $Q$. Despite all of these problems, we were able to extract reliable independent numerical data for the R\'enyi SREE at large values of $\ell$ and small $q$, finding convincing agreement with our TBA computations and yielding an independent confirmation of our method. An example is shown in Fig.~\ref{fig:sree_thermal}(b). 

\subsection{Quantum quenches from tilted Néel states}

In this section we finally consider the GGE reached at late times after a quantum quench. We focus in particular on the tilted N\'eel initial state
\begin{align} \label{eq:tilted_Neel} 
	|\vartheta;\swarrow\nearrow\rangle
	&=
	\left(
	\left[-\sin\left(\frac{\vartheta}{2}\right)|\uparrow\rangle+\cos\left(\frac{\vartheta}{2}\right)|\downarrow\rangle\right]
	\otimes
	\left[\cos\left(\frac{\vartheta}{2}\right)|\uparrow\rangle+\sin\left(\frac{\vartheta}{2}\right)|\downarrow\rangle\right]
	\right)^{\otimes L/2}
	\,, 
\end{align}
where we denoted by $\ket{\uparrow}$, $\ket{\downarrow}$ the two basis states for the local Hilbert space. We are interested in the reduced density matrix reached at large times after the system is initialized in the state~\eqref{eq:tilted_Neel}. Note that our method straightforwardly applies to other quantum quenches from more general integrable initial states~\cite{piroli2017what,pozsgay2019integrable}. More general classes of states can be considered for free models \cite{bc-18,bc-20b,btc-18}.

Importantly, although the titled N\'eel state breaks the $U(1)$ symmetry of the Hamiltonian, the latter is restored at late times during the dynamics, and the GGE density matrix commutes with the magnetization operator. The rapidity distribution function of the GGE associated with the N\'eel ($\vartheta=0$) and tilted N\'eel states were first derived in Refs.~\cite{wouters2014,pozsgay2014correlations} and \cite{piroli2016exact,piroli2017from} respectively, see also~\cite{pozsgay2018overlaps}. Building upon these references, the explicit form of the GGE was extracted from such rapidity distributions in Ref.~\cite{alba2017quench,mestyan2018renyi}. As a result of these analyses, it was found that, for the tilted N\'eel GGE, the functions~\eqref{eq:gn_functions} read $g_n(\lambda)= g_{{\rm N},n}(\lambda) +  g_{\zeta,n}(\lambda)$, where
\begin{align}
	g_{{\rm N},n}(\lambda) &= 
	2 n \ln 4 + 
	\sum_{l=0}^{n-1} \ln \left(
	\frac
	{ \sin^2\left(2 \lambda \right) + \sinh^2\left( \eta (n-1-2l) \right)}
	{4 \tan\left(  \lambda + i \frac{\eta}{2}(n-2 l) \right)\tan\left(  \lambda - i \frac{\eta}{2}(n-2 l) \right)}
	\right)\,,
	\\
	g_{\zeta,n}(\lambda) &= 
	-2 \sum_{l=0}^{n-1} \ln \left[ 2( \cos(2 \lambda) + \cosh( {\eta}(n-1-2l) + 2 \zeta  ) ) \right]\,.
\end{align}

\begin{figure}
\begin{tabular}{lll}
	 \includegraphics[width=0.45\textwidth]{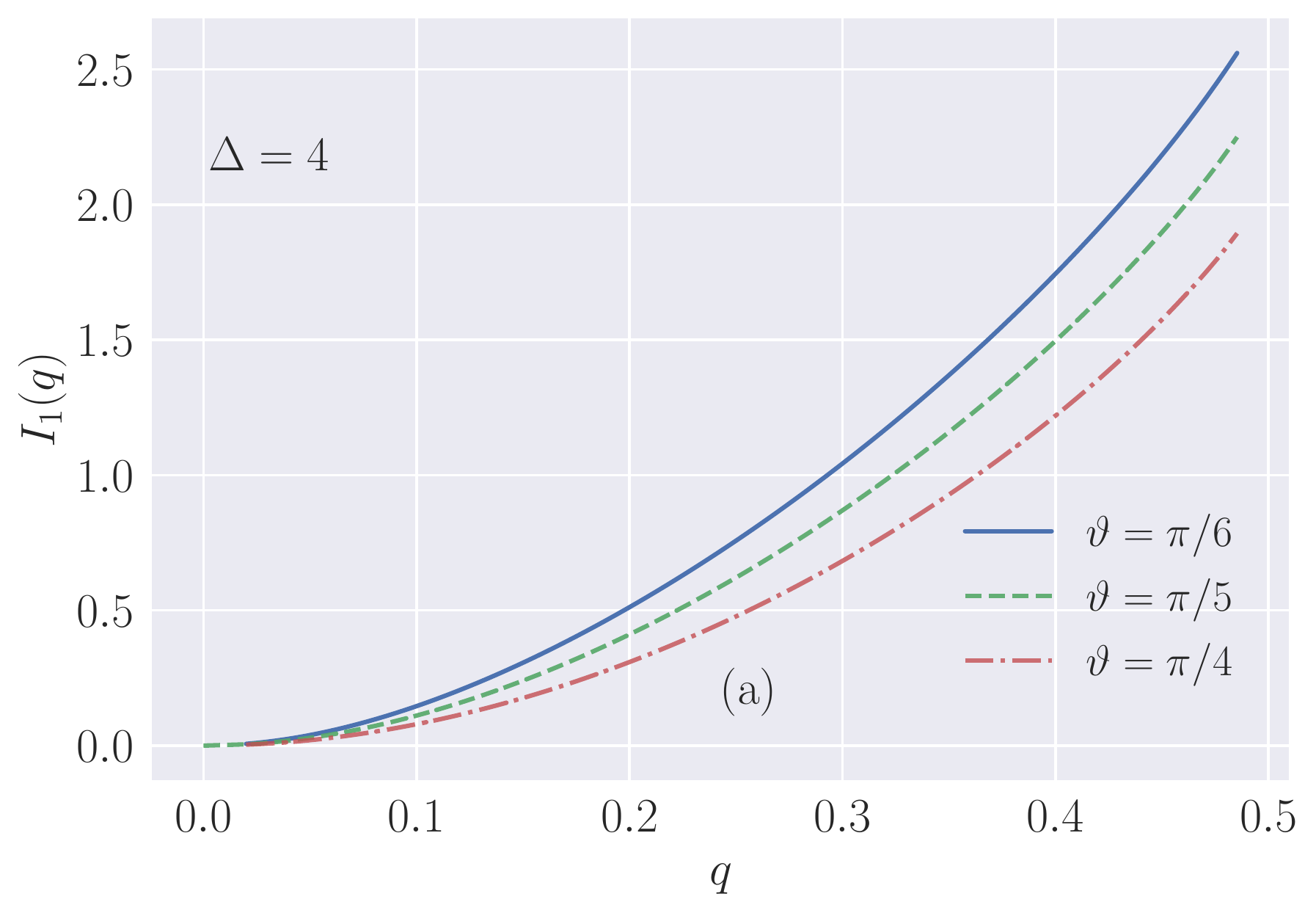}&\qquad \includegraphics[width=0.45\textwidth]{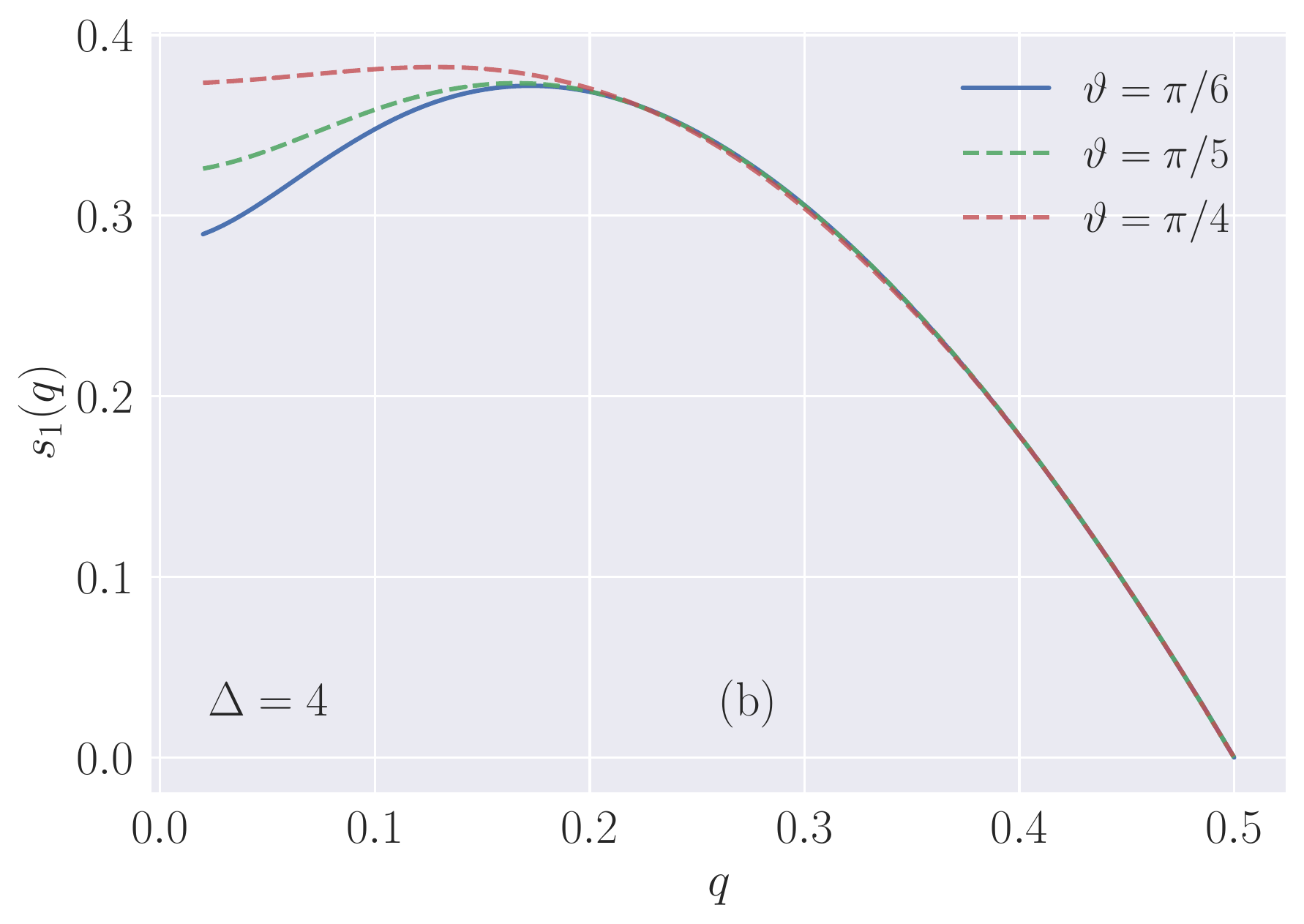}
\end{tabular}
\caption{Rate function $I_1(q)$ (left) and density of symmetry resolved von Neumann entanglement entropy (right) for the tilted N\'eel state, for different tilting angles $\vartheta$. The plots show our results for $q>0$, the curves being manifestly symmetric for $q<0$. }
\label{fig:sree_tiltedNeel}
\end{figure}

We have solved the corresponding TBA equations and evaluated Eq.~\eqref{eq:main_result} to compute the von Neumann SREE. An example of our data is shown in Fig.~\ref{fig:sree_tiltedNeel}, where we also report our results for the rate function $I_{1}(q)$ defined in~\eqref{eq:rate_function_alpha}. Similarly to the thermal case, we observe a generic non-monotonic behavior for the SREE. In addition, we see that the latter exhibits very weak dependence on the tilting angle $\vartheta$.
We have verified that, as the tilting angle $\vartheta$ approaches zero, the rate function grows to infinity, signaling the fact that the GGE density matrix associated with the N\'eel state only has non-trivial support on the sector of vanishing magnetization. Conversely, we found that the von Neumann SREE approaches a non-trivial finite value as $\vartheta\to 0$. 

In this case, we were not able to test our results against independent numerical calculations based on the iTEBD algorithm. While sufficiently large subsystem sizes $\ell$ are needed in order to be able to neglect finite-size effects, the relaxation times are expected to grow at least linearly with $\ell$. On the other hand, the iTEBD algorithm is limited to relatively short times, and does not allow us to explore the late-time regime beyond intervals of a few sites.

\section{Conclusions}
\label{sec:conclusions}

In this work, we have developed a general approach to compute the SREE of thermodynamic macrostates in interacting integrable systems, which is based on a combination of the TBA and the G\"artner-Ellis theorem from large deviation theory. As our main result, we have shown that the von Neumann SREE coincides with the Yang-Yang entropy of an effective macrostate which is determined by the charge sector, see Eq. \eqref{eq:main_result}. 
Although our results apply to any Bethe ansatz integrable model, for concreteness we have presented explicit results only in the case of the XXZ Heisenberg spin chain, focusing on equilibrium thermal states and GGEs following a quench from tilted N\'eel initial states (the former have also been tested against numerical simulations). 
Besides their interest per se, our findings represent a first step towards the derivation of a semiclassical formula for the dynamics of the SREE in interacting integrable systems, thus generalizing the recent results obtained for free fermions in Refs.~\cite{parez2021exact,parez2021quasiparticle}.
Furthermore, we expect them to also be the starting point for the entanglement resolution in non-homogeneous settings in the context of 
generalized hydrodynamics \cite{Bertini2016,CastroAlvaredo2016}, thus extending known results for the total entropies \cite{Al-18,bertini-2018a,abf-19,alba2021generalized}.

\section{Acknowledgments}

Pasquale Calabrese acknowledges support from ERC under Consolidator grant number 771536
(NEMO).

\end{document}